\documentclass[12pt]{iopart}

\usepackage{amsfonts, bm, bbm, graphicx, epsfig, epstopdf, dcolumn, textcomp, soul}
\begin{document}

\title{Single-photon source based on Rydberg exciton blockade}

\author{Mohammadsadegh Khazali}
\address{Institute for Quantum Science and Technology and Department of Physics
and Astronomy, University of Calgary, Calgary T2N 1N4, Alberta, Canada}

\author{Khabat Heshami}
\address{National Research Council of Canada, 100 Sussex Drive, Ottawa, Ontario K1A~0R6, Canada}

\author{Christoph Simon}
\address{Institute for Quantum Science and Technology and Department of Physics
and Astronomy, University of Calgary, Calgary T2N 1N4, Alberta, Canada}

\vspace{10pt}
\begin{indented}
\item[]February 2017
\end{indented}

\begin{abstract}
We propose to implement a new kind of solid-state single-photon source based on the recently observed Rydberg blockade effect for excitons in cuprous oxide. The strong interaction between excitons in levels with high principal quantum numbers prevents the creation of more than one exciton in a small crystal. The resulting effective two-level system is a good single-photon source. Our quantitative estimates suggest that GHz rates and values of the second-order correlation function $g_2(0)$ below the percent level should be simultaneously achievable.
\end{abstract}

\section{Introduction}

\begin{figure} [t]
\centering
\scalebox{0.6}{\includegraphics*[viewport=0 60 490 340]{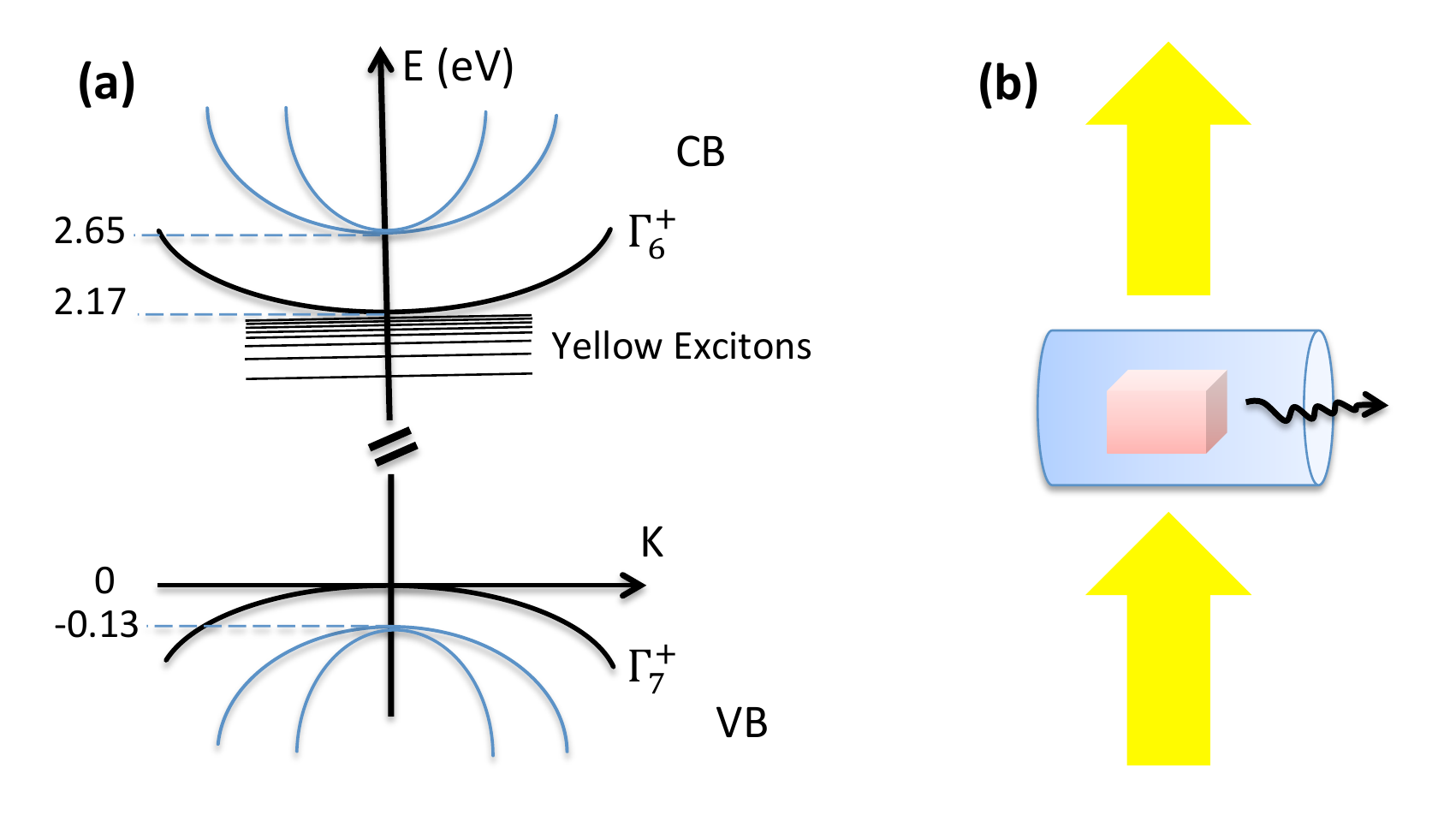}}
\caption{ (Color online) Principle of our proposal. (a) Level scheme of the yellow Rydberg excitons. The $nP$ yellow Rydberg excitons are formed from the $\Gamma_7^+$ valence, $\Gamma_6^+$ conduction and  $\Gamma_4^-$ envelope bands, resulting in a symmetric wavefunction called ortho-exciton. These excitons can be excited via a single-photon dipole transition. (b) Our single-photon source proposal is based on the optical generation of a Rydberg exciton in a crystal that is smaller than the blockade volume. The level shift due to the strong interaction between Rydberg excitons prevents the creation of more than a single exciton, resulting in the formation of an effective two-level system, which can serve as a single-photon source. The fluorescence from the single exciton can be spatially separated from the pump laser by embedding the crystal in a waveguide or cavity in order to select
(and enhance, in the case of a cavity) the emission transverse to the pump.} \label{Scheme}
\end{figure}

Rydberg atoms, i.e. atoms that are excited to high principal numbers $n$, have revolutionized atomic physics and resulted in a variety of applications based on their long lifetimes and strong interactions, which are both due to the large extent of their electronic wave functions. Rydberg atoms have been used in quantum information applications such as atomic \cite{Saf10} and photonic \cite{PhGate1, PhGate2, PhGate3} quantum computation, as well as for single-photon sources \cite{Saf02,Mul13,Dud12, Bar12}.  Furthermore Rydberg atoms have been used in the study of strongly correlated plasmas \cite{plas}, in quantum non-linear optics \cite{NonLinear1, NonLinear2}, in ultracold chemistry (e.g. Rydberg molecules) \cite{Ben09}, and in the study of exotic quantum phases \cite{Pup10,Hen10} and many-body entangled states \cite{Kha16,Gil14}.

In the solid state, hydrogen-like series of excited states have also been observed in the semiconductors cuprous oxide ($Cu_2O$) \cite{Gro56, Bayer} and $WS_2$ \cite{Che14}.
Excitons are optically excited electron-hole pairs in a semiconductor that are bound by the Coulomb interaction in analogy with atoms. A recent experiment \cite{Bayer} demonstrated the phenomenon of Rydberg blockade for excitons in cuprous oxide, where the presence of a highly excited exciton prevents the creation of more excitons in its vicinity by shifting the corresponding states out of resonance with the exciting field because of the strong interaction between Rydberg excitons. Quantum coherent effects for these excitons have also recently been demonstrated \cite{Scheel}.

The discovery of Rydberg states in solid-state systems is promising for the development of scalable and integrated quantum devices. As a first step in this direction, we here study the possibility of realizing a single-photon source based on Rydberg exciton blockade in cuprous oxide. Single-photon sources are useful for  quantum cryptography \cite{BB84},  quantum repeaters \cite{San07}, and photonic quantum computing \cite{KLM}. Single-photon sources have been proposed and realized in different systems \cite{approach}, including quantum dots \cite{Pel02,kei04,Clo10} and Rydberg atoms \cite{Saf02,Mul13,Dud12, Bar12}.

Our Rydberg exciton single-photon source proposal is based on pumping a cuprous oxide crystal that is smaller than the blockade volume by a monochromatic, continuous-wave laser in order to generate yellow $nP$ Rydberg excitons (see Fig. 1a). Under these conditions, the level shift due to the strong interaction between Rydberg excitons prevents the creation of more than a single exciton, resulting in the formation of an effective two-level system, which can serve as a single-photon source. The single-photon emission can be spatially separated from the pump laser by enhancing the emission in a direction transverse to the pump via a cavity using the Purcell effect \cite{Purcell} or by embedding the crystal within a waveguide (see Fig. 1b).

This paper is organized as follows. We begin with an introduction about Rydberg excitons in Sec.~\ref{Sec:exciton}.
In Sec.~\ref{Sec:Blockade} we define the blockade condition for different driving field regimes and also discuss collective enhancement.
We then explain our scheme in Sec.~\ref{Sec:scheme}, followed by a discussion on optimizing the photon generation rate in Sec.~\ref{Sec:Optimum}.
In Sec.~\ref{Sec:EfficiencyBlockade} we study the persistence of the blockade effect for different crystal sizes and over a wide range of driving intensities by quantifying the two-exciton generation probability and its effect on the zero-time second-order correlation function $g_2(0)$. In Sec.~\ref{Sec:ExperimentalRealization} we propose the specifications and parameters for an experimental realization with excellent performance.
In Sec.~\ref{Sec:Outlook} we discuss our results and look towards the future.

\section{Rydberg Excitons}
\label{Sec:exciton}

An incident photon can excite an electron in a semiconductor to the
conduction band, leaving a hole in the valence band. This electron-hole pair may be
bound by the Coulomb interaction to form an integer-spin particle, called an exciton. This
system can be modeled as a hydrogen-like atom in which the relative motion of the electron
and the hole can be characterized by the exciton radius which is proportional to the
square of the principle quantum number. For the cuprous oxide excitonic system with electron charge $e$, dielectric constant $\epsilon=7.5$, and reduced mass $\mu$, the Bohr radius is given by $a_B=\frac{\hbar^2}{\mu (e^2/\epsilon)}=1.1nm$ \cite{Kav97}.
The average radius of the Rydberg exciton is given by \cite{Galagher}
\begin{equation}
\langle r_n \rangle=\frac{a_B}{2} (3n^2-l(l+1)),
\label{rn}
\end{equation}
where $n$ is the principal quantum
number and $l$ is the angular momentum. The average radius for a Rydberg exciton with $n=25$ is 1$\mu m$, which corresponds to an atomic radius for $n=140$. The 2$\mu m$ effective size of the mentioned state
is about ten times larger than the photon wavelength in the crystal (which is 211 nm). This fact has raised concerns about the validity of dipole approximation \cite{Bayer}. However, one should note that the transition dipole moment $d=\langle g|r|e\rangle$ in this case is between the spatially compact non-excitonic state $|g\rangle$ and the vastly extended Rydberg exciton state $|e\rangle$. The dipole approximation should still be valid in this case due to the fact that the external field is constant over the range that the transition dipole moment  is non-zero, which is where the two wave-functions have a significant overlap. The relevant length scale is therefore set by the dimension of the non-excitonic state.

The energy of Rydberg excitons is given by
\begin{equation}
E_n=E_g-\frac{Ry}{(n-\delta_l)^2},
\end{equation}
where $Ry=92meV$ is the Rydberg constant, $E_g$ is the exciton's bandgap. The latter can be in the yellow, green, blue, or violet parts of the spectrum, depending on the combination of valence and conduction bands. The deviation from hydrogen-like behavior is quantified by the quantum defect $\delta_l$ that depends on the envelope bands $l=S,P,D,...$ as measured in Ref.~\cite{Quantum defects,Quantum defects2}. Here we follow Ref.~\cite{Bayer} in considering yellow $P$ excitons with  $E_g=2.17208eV$ and $\delta_p=0.23$.

\section{Blockade in Two Regimes}
\label{Sec:Blockade}

We propose to excite the crystal with a continuous wave laser to generate Rydberg excitons. An important consequence of the Rydberg interaction is a position-dependent level shift, which prevents the generation of multiple Rydberg excitons within the blockade volume, by making the transition out of resonance with the exciting field.

In the strong field regime, where the collective Rabi frequency is much larger than other line broadening mechanisms, the blockade condition is given by
\begin{equation}
\sqrt{N}\Omega \approx \frac{C_3}{R_b^3}.
\label{Blockade}
\end{equation}
Here we take $N=\frac{V}{\langle r_n \rangle^3}$ to be the number of excitons that can be fitted within either the blockade or the crystal volume, whichever is smaller ($V=$min$[V_c,V_B]$),  and $\langle r_n \rangle$ is the average radius of the Rydberg excitons given by Eq.~\ref{rn}. This choice of $N$ and the corresponding definition of the collective Rabi frequency $\Omega'=\sqrt{N}\Omega$ is based on the experimental observation and theoretical description of superradiance for excitons that are weakly coupled to quantum dots \cite{superradiance}, in combination with the observation of collective excitation in atomic Rydberg experiments \cite{Collective}. In Ref. \cite{superradiance} it was found that the correct choice of $N$ for describing the superradiance is the number of excitons that can be fitted within the quantum dot, in the regime where the volume of the quantum dot is significantly larger than the size of exciton. In comparison, in Rydberg atom experiments $N$ is given by the number of atoms within the blockade volume.
Eq. \ref{Blockade} has been verified over a wide range of intensities and densities in terms of the dimensionless parameter of $\alpha=\frac{\hbar \Omega}{C_6 \rho^2}$, namely $10^{-9}<\alpha<10^{-1}$ \cite{high-density,high-density1,high-density2}, justifying its application for our proposal where $\alpha \approx 0.1$.

In the weak field regime $\Omega' \ll \Gamma_n$ the blockade is defined by comparing the interaction-induced energy shift to the linewidth $\Gamma_n$. Measurement of the blockade volume for the yellow $P$ excitons \cite{Bayer} confirms the expected scaling of
\begin{equation}
V_{blockade}=\frac{4\pi}{3}\frac{C_3}{\Gamma_n/2}=3.10^{-7}\mu m^3 \times n^7.
\end{equation}
Blockade volumes as large as 2000$\mu m^3$ have been observed for $n=24$.

\section{Scheme dynamics}
\label{Sec:scheme}

In the case that the crystal is smaller than the blockade volume it can be considered as a two level system. The fluorescence emissions from this two level system are in the form of single photons, see below. To separate the generated single photons from the strong pump beam we propose to confine the crystal within a cavity. Due to the Purcell effect the single photon emission is enhanced along the direction of the cavity, which can be perpendicular to the pump laser, see figure \ref{Scheme}. A moderate Purcell effect is sufficient. In fact, large Purcell factors are undesirable because they would lead to mixing between different Rydberg levels (because the Purcell effect increases the linewidth), which could destroy the blockade. Although currently natural crystals are more pure than the synthesized ones, growing the crystal \cite{growing} between the cavity mirrors would be one way to avoid optical losses due to unwanted interfaces. Another potential method for separating the generated single photons is by embedding the crystal within a hollow-core optical fiber while shining the laser perpendicular to the fiber, see figure \ref{Scheme}b.

The time evolution of the effective two-level system  with natural frequency $\omega_0$, excited with a monochromatic laser field $E=E_0 \, e^{-i\omega t}$, is given by the master equation
\begin{eqnarray}\label{Master}
&&\dot{\rho}_{ee}=-\Gamma \rho_{ee} + i\frac{\Omega'}{2} (\rho_{eg} -\rho_{ge}) \\ \nonumber
&&\dot{\rho}_{eg}=i\frac{\Omega'}{2} (\rho_{ee}-\rho_{gg})+ (-\frac{\Gamma}{2}+i \Delta \omega) \rho_{eg},
\end{eqnarray}
where  $\Gamma$ is the spontaneous emission rate, $\Delta \omega=\omega-\omega_0$ is the detuning of the laser from the transition,  $\rho_{gg}=1-\rho_{ee}$ and $\rho_{eg}=\rho_{ge}^*$. The steady-state solution for the exciton population is
\begin{equation}
\rho_{ee\, (t\rightarrow \infty)}(\Delta\omega)=\frac{\frac{\Omega'^2}{4}}{\Delta \omega^2+\frac{\Gamma^2}{4}(1+2\frac{\Omega'^2}{\Gamma^2})},
\label{rho11}
\end{equation}
which exhibits a Lorantzian linewidth with a power broadened full-width at half maximum of $FWHM=\Gamma\sqrt{(1+2\Omega'^2/\Gamma^2)}$. A more accurate vision of the spectrum of the emitted photons can be obtained from the Fourier transform of the first-order correlation function. In the strong driving field regime the spectrum contains three peaks known as the Mollow triplet \cite{Mollow}, centered at $\omega_0$ and $\omega_0\pm \Omega$ with widths of $\Gamma$ and $3\Gamma/2$ respectively. Applying the regression theorem \cite{Scully}, the second-order correlation function is
\begin{eqnarray*}
g^{(2)}(\tau)&=& \rho_{ee}^{-1}(t\rightarrow \infty) \, \left.\rho_{ee}(\tau)\right|_{\rho(0)=|g\rangle\langle g|}   \\ \nonumber
&=&1-e^{-\frac{3\Gamma}{4}\tau}(\cos(\Omega_e \tau)+\frac{3\Gamma}{4\Omega_e} \sin(\Omega_e \tau)),
\label{g2}
\end{eqnarray*}
where $\Omega_e=\sqrt{\Omega'^2-\frac{\Gamma^2}{16}}$ is the effective Rabi frequency.  A value of the zero-time correlation function of $g_2(0)=0$ corresponds to perfect single-photon character of the emission.

\section{Optimum Regime}

\begin{figure} [t]
\centering
\scalebox{0.45}{\includegraphics*[viewport=5 250 700 555]{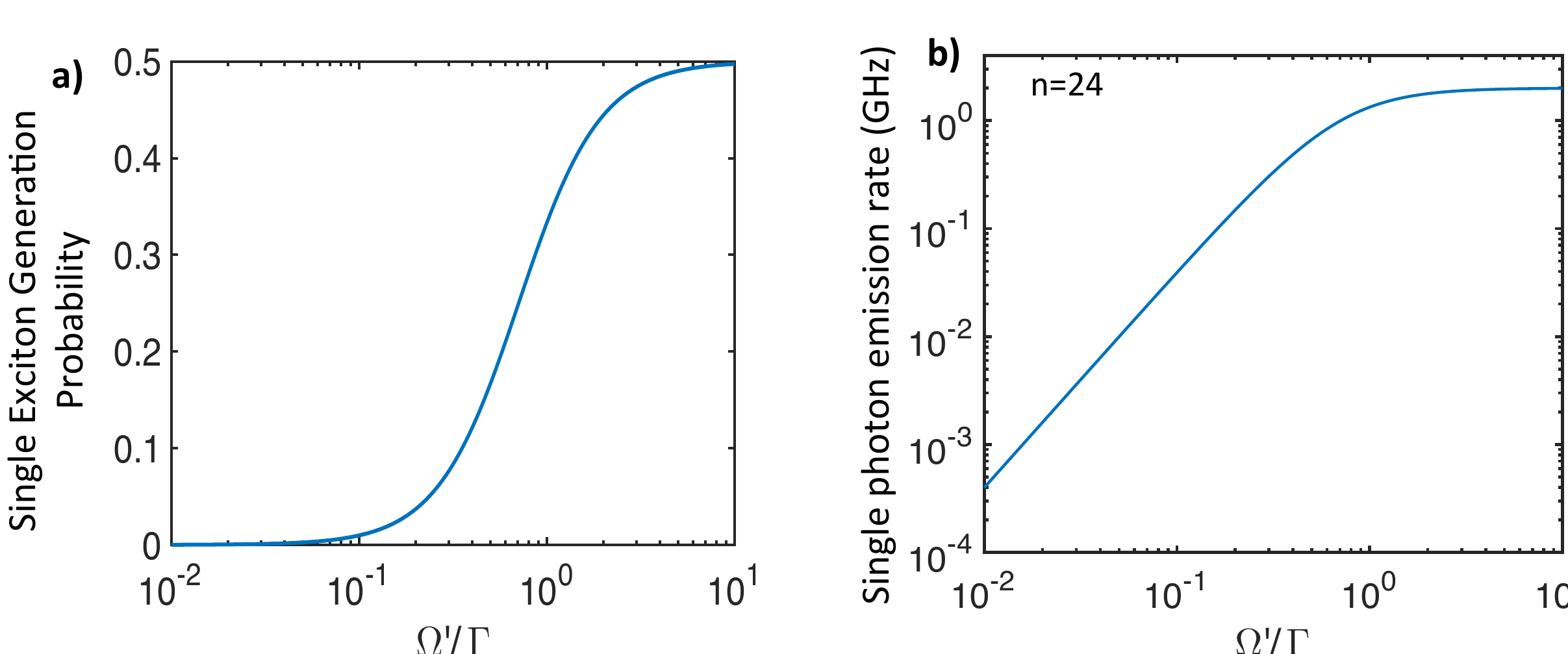}}
\caption{ (Color online)  (a) The excited state population $\rho_{ee}$ increases with the driving field amplitude. (b) As a consequence, the single-photon generation rate $\rho_{ee}\Gamma$ (here shown for the example of $n=24$) also increases.} \label{intensity}
\end{figure}

\label{Sec:Optimum}
Figure~\ref{intensity} represents the probability of exciton generation (Eq.~\ref{rho11}) and a case study of the single-photon generation rate for $24P$ excitons, both as a function of driving strength $\frac{\Omega'}{\Gamma}$. This shows that the strong field regime $\frac{\Omega'}{\Gamma} \gg 1$ allows higher rates. A larger crystal volume $V$ enhances the collective Rabi frequency $\Omega'=\sqrt{\frac{V}{\langle r_n \rangle^3}}\Omega$, but $V$ has to be small enough to remain in the blockade regime, see below. To ensure the integrity of the exciton and the validity of our approximation for $N$, it is also important that the crystal is significantly larger than the exciton size.

Regarding the exciton principal number $n$, while higher $n$ reduces the spontaneous emission rate $\Gamma_n\propto d_n^2 \propto n^{-3}$, the effective Rabi frequency also decreases ($\Omega \propto d_n \propto n^{-3/2}$), while the quantities determining the collectivity scale as $\langle r_n \rangle \propto n^2$ and $R_b \propto n^{1.5}$, yielding an effective scaling of the collective Rabi frequency $\frac{\Omega'}{\Gamma}\propto n^{3/4}$, suggesting that high principal numbers are advantageous. Higher principal numbers also help to simultaneously fulfill the strong blockade and weak exciton confinement conditions, since the ratio of exciton size to blockade volume scales with $n^{-1}$. Experimentally Rydberg blockade with the expected scaling behavior has been demonstrated in the presence of background noise for principal numbers as high as $n=24$ \cite{Bayer}.

\section{Effectiveness of Blockade and Second-Order Correlation Function}
\label{Sec:EfficiencyBlockade}

\begin{figure} [t]
\centering
\scalebox{0.45}{\includegraphics*[viewport=5 30 700 340]{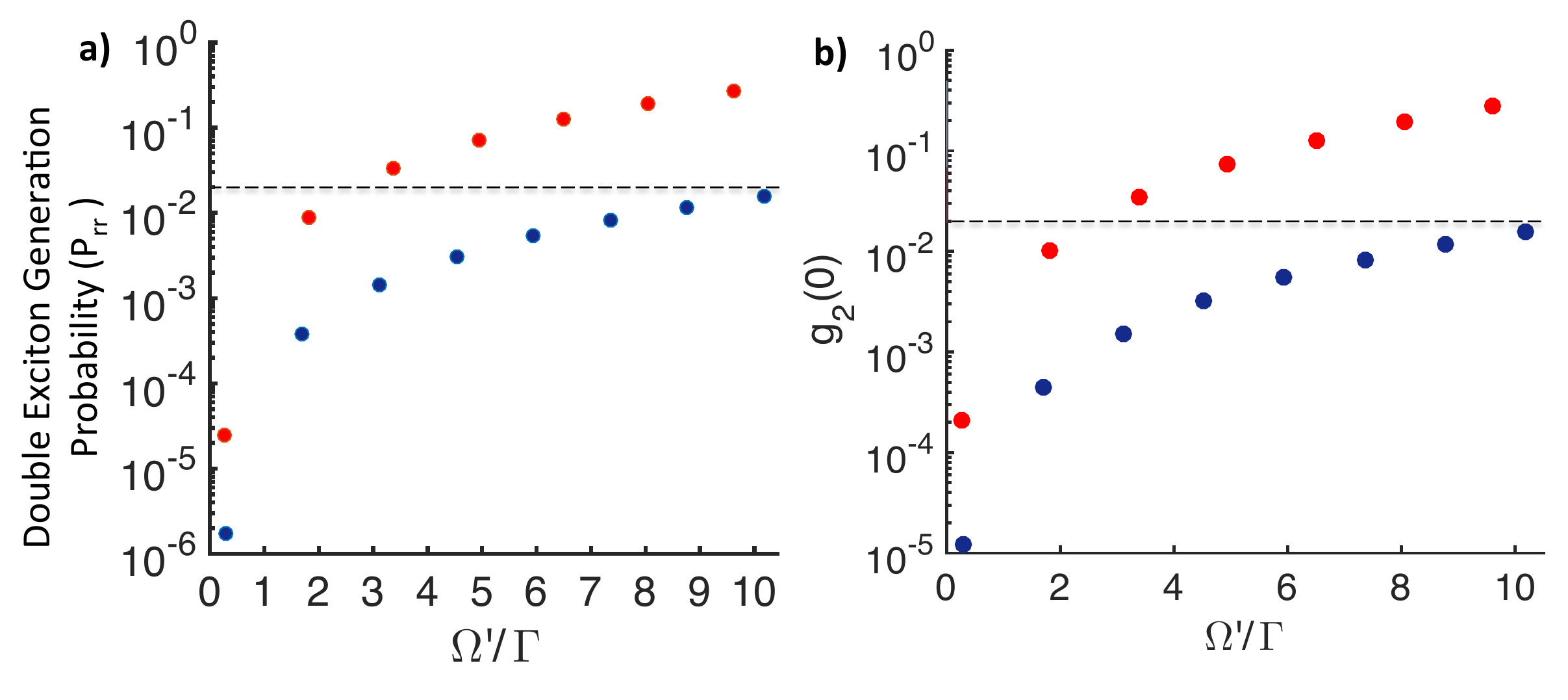}}
\caption{ (Color online) Effectiveness of Rydberg blockade as a function of driving field amplitude for different crystal sizes. (a) Two-exciton generation probability and (b) second-order correlation function $g_2(0)$, for principal number $n=24$ and two different cubic crystals with sides of length $4\mu m$ (blue circles) and $6\mu m$ (red circles).} \label{Fig:blockade}
\end{figure}

We now quantify the extent to which the blockade is effective and the consequences for the second-order correlation function. To calculate the average probability of generating more than one exciton, we consider a situation where the blockade is largely effective, such that at most two Rydberg excitons can simultaneously be excited. The relevant Hilbert space is spanned by the ground state $|0\rangle=|g_1...g_N \rangle$, the single-exciton state $|1\rangle=\sum |g_1...r_i...g_N\rangle$, and the two-exciton states at sites $i$ and $j$,  $|2_{ij}\rangle=|g_1...r_i...r_j...g_N\rangle$. We assume that adjacent exciton sites are separated by $2\langle r_n \rangle$, inspired by the exciton superradiance theory of Ref. \cite{superradiance}, see also Sec. 3.
In the regime that $\rho_{_{2_{ij}2_{ij}}}\ll 1$, one can calculate the double-excitation probability as a perturbation to the steady state solution of Eq.~\ref{rho11}. The master equation reads (omitting the $(ij)$ indices for simplicity)
\begin{eqnarray}\label{MasterBlockade}
&&\dot{\rho}_{22}=-\Gamma \rho_{22} - i\frac{\Omega}{2\sqrt{N}} (\rho_{12}-\rho_{21})  \\ \nonumber
&&\dot{\rho}_{12}=-i\frac{\Omega}{2\sqrt{N}} (\rho_{22}-\rho_{11})+ (-\Gamma/2+i V_{ij}) \rho_{12},
\end{eqnarray}
where $V_{ij}=\frac{C_3}{r_{ij}^3}$ is the dipole-dipole interaction between the two excitons. The steady state solution for the two-exciton generation at sites $(i$ and $j)$ is
\begin{equation}
\rho_{_{2_{ij}2_{ij}}}=\frac{X}{1+2X}\frac{Y}{1+2Y},
\end{equation}
with $X=\frac{N \Omega^2}{\Gamma^2}$ and $Y=\frac{\Omega^2/N}{V_{ij}^2+\Gamma^2/4}$. The total two-exciton generation  probability can be found as $P_{rr}=\sum \limits_{i<j} \rho_{_{2_{ij}2_{ij}}}$. Figure~\ref{Fig:blockade}b  shows the two-exciton generation probability  as a function of different field amplitudes $\frac{\Omega'}{\Gamma}$ for two different cubic crystals with sides of length of $6\mu m$ (red circles) and $4\mu m$   (blue circles) for the $24P$ Rydberg excitons. For the smaller crystal the blockade regime is preserved up to the desired strong field regime of $\frac{\Omega'}{\Gamma}\approx 10$ where the single-photon generation rate is optimal.

For two two-level systems, the second-order correlation function $g_2(0)$ takes a value of  $1/2$, instead of 0 for a single emitter \cite{Mandel}. As a result, the effect of double excitations due to imperfect blockade on the correlation function can be quantified as $g_2(0)=\frac{1}{2} \frac{P_{rr}}{\rho_{ee}}$.   Figure \ref{Fig:blockade}b shows $g_2(0)$ for the mentioned two crystal sizes as a function of the field amplitude.

\section{Proposed Experimental Realization and Expected Performance}
\label{Sec:ExperimentalRealization}

Here we consider the largest exciton principal number $n=24$ that has experimentally been excited \cite{Bayer}. For the laser intensity of 4$\frac{\mu W}{mm^2}$ \cite{Scheel} the Rabi frequency of the transition to $24P$ is $\Omega\approx 9$GHz. The spontaneous emission rate of the yellow exciton $P$ levels has been experimentally measured as $\Gamma(n)=28.n^{-3}$ THz \cite{Bayer} where $\Gamma(24)=2$ GHz. In order to ensure an effective blockade, we consider a cubic crystal with sides of length  $4\mu m$. For the cavity Purcell factor of $2$, these values put the system into the strong field regime, $\frac{\Omega'}{\Gamma}=6$, allowing a high single-exciton generation rate (see Fig.~\ref{intensity}a) while preserving the Rydberg blockade (see Fig.~\ref{Fig:blockade}a). The expected single-photon generation rate for these parameter values is 2$GHz$, and the second-order correlation function  $g_2(0)=0.007$ (Fig.~\ref{intensity}b), corresponding to an excellent single-photon source.

\section{Discussion and Outlook}
\label{Sec:Outlook}

Besides the obvious difference of being in the solid state, one important difference of the present system compared to atomic Rydberg systems is the fact that the Rydberg excitons can be generated by a single-photon transition, whereas in atomic systems two-photon transitions are typically required. On the one hand, the direct single-photon excitation allows a conceptually simpler approach to single-photon generation; on the other hand it makes it more challenging to filter out the pump (here we rely on spatial filtering).
In addition to promising scalability and integration, our proposal allows for higher rates compared to atomic systems, mainly because high Rabi frequencies are difficult to achieve for high principal numbers in atomic Rydberg systems.

Our proposal differs from other solid-state approaches to single-photon sources such as quantum dots or individual defects (e.g. nitrogen-vacancy or silicon-vacancy centers in diamond) by the fact that the emitter is significantly larger ($\mu$m instead of nm scale). It is an interesting question how this difference in size impacts common challenges for solid-state single-photon sources, in particular spectral diffusion. Taking an optimistic perspective, the larger volume might help to average out effects due to fluctuating charges, making the system more robust to such fluctuations.

We have focused on continuous-wave excitation. Driving the proposed system with a pulsed pump should allow the creation of single photons on demand, and driving it with two phase-coherent pulses might even allow the generation of photonic time-bin qubits. The system needs to be in the regime where the pump is spectrally broader than the Rydberg radiative linewidth and narrower than the level spacing with respect to the nearest Rydberg level. Phase-coherent pump pulses have been used for the generation of time-bin entangled photon pairs through biexciton-exciton cascade in quantum dots~\cite{Jayakumar14}.

As mentioned in the introduction, the generation of single photons is only the first step towards realizing the full potential of Rydberg excitons for quantum photonics. For example, the present approach may also be suitable for Rydberg-exciton based single-photon subtraction \cite{Hon11,kub13} schemes, with applications to photon counting and photonic quantum state generation. Another interesting avenue is the generation of photonic quantum gates inspired by Rydberg-atom proposals \cite{PhGate1,PhGate2,PhGate3}.

\vspace{0.5cm}

{\it Acknowledgments.} We acknowledge financial support from NSERC and thank S. Goswami for fruitful discussions.

\end{document}